%% file: ieee_revision1.tex
\documentclass[lettersize,journal]{IEEEtran}
\usepackage{amsmath,amsfonts}
\usepackage{algorithmic}
\usepackage{algorithm}
\usepackage{array}
\usepackage[caption=false,font=normalsize,labelfont=sf,textfont=sf]{subfig}
\usepackage{textcomp}
\usepackage{stfloats}
\usepackage{url}
\usepackage{verbatim}
\usepackage{graphicx}
\usepackage{cite}
\usepackage{etoolbox}
\newtoggle{finalversion}

\toggletrue{finalversion}

\newcommand{\annot}[1]{%
\iftoggle{finalversion}{%
#1
}
{\textcolor{blue}{#1}}
}

\usepackage{hyperref}
\usepackage[square,sort,comma,numbers]{natbib}
\input{new_math}
\newcommand{\dip}[1]{implicit generative prior}
\newcommand{\Dip}[1]{Implicit generative prior}

\begin{document}

\title{Low-light phase retrieval with \dip{}s}

\author{Raunak Manekar*, Elisa Negrini*, Minh Pham, Daniel Jacobs, Jaideep Srivastava, Stanley J. Osher, Jianwei Miao
\thanks{*these authors contributed equally to this work. \textit{corresponding author: R. Manekar (manek009@umn.edu)} R. Manekar and J. Srivastava are with the Department of Computer Science and Engineering, University of Minnesota, Minneapolis, MN, USA. (email:  manek009@umn.edu, srivasta@umn.edu. S.J.Osher and E. Negrini are with the Department of Mathematics, University of California, Los Angeles, Los Angeles, CA, USA (email: sjo@math.ucla.edu, enegrini@math.ucla.edu). J. Miao, M. Pham and D. Jacobs are with the Department of Physics and Astronomy, and California NanoSystems Institute, University of California,
Los Angeles, CA, USA (email: j.miao@ucla.edu, minhrose@ucla.edu, danieljacobs@physics.ucla.edu)}
}

\markboth{Journal of \LaTeX\ Class Files,~Vol.~14, No.~8, August~2021}%
{Shell \MakeLowercase{\textit{et al.}}: A Sample Article Using IEEEtran.cls for IEEE Journals}


\maketitle
  
\begin{abstract}
Phase retrieval (PR) is fundamentally important in scientific imaging and is crucial for nanoscale techniques like coherent diffractive imaging (CDI). Low radiation dose imaging is essential for applications involving radiation-sensitive samples. However, most PR methods struggle in low-dose scenarios due to high shot noise. Recent advancements in optical data acquisition setups, such as in-situ CDI, have shown promise for low-dose imaging, but they rely on a time series of measurements, making them unsuitable for single-image applications. Similarly, data-driven phase retrieval techniques are not easily adaptable to data-scarce situations. Zero-shot deep learning methods based on pre-trained and implicit generative priors have been effective in various imaging tasks but have shown limited success in PR. In this work, we propose low-dose deep image prior (LoDIP), which combines in-situ CDI with the power of implicit generative priors to address single-image low-dose phase retrieval. Quantitative evaluations demonstrate LoDIP’s superior performance in this task and its applicability to real experimental scenarios.
\end{abstract}


\section{Introduction}

\input{intro}

\section{Related Work}
\input{m_related_work}

\section{Proposed method}
\input{method}

\section{Experimental Results and Discussion}
\input{m_experiments}

\section{Conclusion and Future Work}
Emerging deep learning methods present several opportunities to improve upon and expand the scope of computational imaging techniques framed as mathematical optimization problems. The LoDIP approach, introduced in this work, integrates the optimization framework of \dip{}s with physical constraints from a modified CDI setup. This combination results in high-SNR reconstructions while preserving resolution at low photon counts in strong presence of noise where traditional methods are known to struggle.
Experiments demonstrate that LoDIP's performance is consistent across different natural images, realistic biological cells and experimental data.
We expect the LoDIP method to find applications in X-ray imaging of dose-sensitive samples across diverse fields including organic semiconductors and biological specimens.

\section*{Acknowledgments}
This work was supported by STROBE, a National Science Foundation Science \& Technology Center, under grant number DMR-1548924, and the U.S. Air Force Office of Scientific Research Multidisciplinary University Research Initiative (MURI) program under Award No. FA9550-23-1-0281. E.N. was partially supported by the Simons Postdoctoral program at IPAM, NSF DMS-1925919, and NSF-2331033. The authors acknowledge the Minnesota Supercomputing Institute (MSI) at the University of Minnesota for providing resources that contributed to the research results reported in this paper.

\footnotesize
\bibliography{references}
\bibliographystyle{IEEEtran}

\end{document}

%% file: new_math.tex
\usepackage{amscd,amssymb,bm,url,color,latexsym}

\allowdisplaybreaks

\usepackage[capitalize,nameinlink]{cleveref}

\renewcommand{\mathbf}{\boldsymbol}

\newcommand{\parahead}[1]{\textbf{#1}}

\newcommand{\mb}{\mathbf}
\newcommand{\mc}{\mathcal}

\newcommand{\bb}{\mathbb}

\newcommand{\Cp}{\bb C}

\newcommand{ \paren }[1]{ \left( #1 \right) }

\newcommand{\abs}[1]{\left| #1 \right|}

\numberwithin{equation}{section}

%% file: intro.tex
Coherent diffractive imaging (CDI) is a lensless imaging technique \cite{miao1999extending} used for high resolution imaging at nanoscale. CDI has found broad applications across different disciplines due to its remarkable ability to provide high-resolution structural information about a wide range of specimens from biological to nanoscale objects \cite{miao2015beyond}. Unlike visible light, X-rays have high penetrating power and thus can be used to image thick, unfixed specimens. However, many samples of interest for CDI, such as biological materials, polymers or organic semiconductors, require minimal radiation exposure to prevent damage during data acquisition \citep{GeorgePhotochem2012,GarmanWeik2017}. Thus, there is considerable interest in techniques that minimize sample radiation exposure and enable X-ray imaging at extremely low photon counts. However, imaging at low photon counts is very challenging as it leads to very high shot noise in the acquired measurements (diffraction pattern) and makes the subsequent image reconstruction (i.e. phase retrieval) very hard.
Under high shot noise, iterative phase retrieval algorithms, including ER, HIO, become unstable, as they often get trapped in local minima, stagnate, and fail to converge \cite{ShechtmanEtAl2015Phase,pham2019generalized}.

\parahead{Motivation.} One category of proposed solutions for the low-dose challenge involve modifications to the optical data acquisition setup of CDI \cite{putkunz2011phase,lan2014method,lo2018situ}. A notable example is in-situ CDI, as introduced by Lo et al. \cite{lo2018situ} and further explored in \cite{lu2023computational}. A dose-tolerant static region is positioned next to the sample and is illuminated with a higher radiation dose. This ensures sufficient light reaches the detector while maintaining a low radiation dose on the sample. Multiple measurements are taken over time and the static region is leveraged as a robust time-overlap constraint which regularizes the phase retrieval optimization.
However, in order to leverage the time-invariant static region as real-space constraint, this method requires multiple measurements (that is multiple diffraction patterns). Adapting this setup for single-image scenarios results in suboptimal reconstructions (see HIO-stat column in \cref{fig:ch3_experimental_results_bio_sample} ).

On the computational front, data-driven supervised learning approaches have shown potential for simpler versions of phase retrieval at low-photon counts \cite{goy2018low,deng2020learning,metzler2021deep}. However, the necessity for a substantial dataset is a challenge to meet for numerous scientific imaging applications. Furthermore, these methods are prone to generalization problems if the training data diverges from the target domain.
Consequently, deep learning-based zero-shot methods such as deep image prior (DIP) \cite{ulyanov2018deep,sitzmann2020implicit} have been employed. These do not require pre-training, rather the primary concept is to leverage the inductive bias induced by the neural network architecture as a prior for natural images. We refer to this class of methods as \dip{}s to highlight the absence of training data.
While successful in simpler phase retrieval scenarios \cite{jagatap2019algorithmic,heckel2018deep,bostan2020deep}, these methods often struggle to converge to satisfactory solutions in more challenging realistic scenarios, such as the single-shot far-field Phase Retrieval (FFPR) studied in this work. This observation is supported by our low-dose experiments (see Fig. \ref{fig:ch3_experimental_results_bio_sample}) and corroborated by prior literature \cite{tayal2020inverse,zhuang2022practical}. The difficulty can be attributed to the well-known trivial ambiguities in the phase retrieval problem \cite{ShechtmanEtAl2015Phase}.

\parahead{Basic Idea.} 
LoDIP, a single-image method for low-dose phase retrieval, modifies both the experimental setup and computational algorithm of classical CDI. We modify the imaging setup of CDI to incorporate a static region illuminated with a high dose of radiation alongside the sample of interest which is illuminated by a low dose. Then we modify the DIP framework to exploit the additional physical constraints coming from this setup. The high-dose static region increases the amount of light incident on the detector, without increasing the radiation dose on the sample. This results in a reduction of noise in the captured diffraction pattern. Moreover, the static region aids in the convergence of \dip{} by alleviating the trivial ambiguities \cite{hyder2020solving}. 

\annot{
\parahead{Innovation.}
LoDIP introduces innovation by concurrently modifying both experimental setup and computational algorithm. By integrating physics-based constraints from the modified experimental setup, LoDIP effectively resolves instability and artifacts typical of conventional DIP methods in single-shot far-field phase retrieval. This ensures stable optimization and precise reconstruction from low-light diffraction patterns, leveraging the benefits of deep image prior optimization while overcoming inherent challenges in single-shot far-field phase retrieval.}

\annot{
Unlike in-situ CDI setup, which requires multiple diffraction patterns, LoDIP operates in a single-image framework. Additionally, in contrast to in-situ CDI, where the static region is used as a time-invariant constraint present in multiple measurements, in LoDIP, the high-dose imaging of the static region serves two novel purposes: (a) 
being imaged at high-dose, it provides an accurate reconstruction of the static region, which acts as a strong constraint crucial for achieving a high-quality reconstruction of the sample; and (b)  enhancing the convergence of phase retrieval optimization by mitigating ambiguities arising from symmetries in the forward process, a significant challenge in phase retrieval \cite{ShechtmanEtAl2015Phase,hyder2020solving}.}


\parahead{Contributions.} Experiments demonstrate that LoDIP is effective for both simulated and experimental data. The LoDIP framework is very flexible, compatible with any \dip{} method, and can be tailored to diverse experimental setups. Unlike iterative phase retrieval algorithms, LoDIP does not necessitate specialized tuning to optimize the algorithmic parameters for satisfactory results, making LoDIP a user-friendly tool for a wide user community. The contributions of this work can be summarized as follows:

\begin{itemize}
    \item A zero-shot method for low-dose phase retrieval. LoDIP uses the inductive bias of an implicit generative prior and does not require large dataset for training.
    \item Extensive experiments on simulated and experimental datasets demonstrate that LoDIP outperforms state-of-the-art methods across a range of image quality metrics.
    \item On the experimental side, LoDIP being designed for single-image phase retrieval, can produce reconstructions comparable to the original in situ CDI method without requiring a time series of measurements.
\end{itemize}

%% file: m_related_work.tex
\subsection{Coherent Diffractive Imaging} \label{cdi_description}

In Coherent Diffractive Imaging (CDI), an object is illuminated by a highly coherent light source. The interaction between the object and the incident wave results in the generation of a diffraction pattern, which is subsequently detected. However, while detectors capture the magnitude of the diffracted wave, the phase information is lost. As a consequence, the process of reconstructing the image of the object of interest necessitates the development of a computational algorithm designed to recover the lost phase from the acquired diffraction pattern. This is commonly referred to as the ``phase retrieval" problem. As shown in \cite{miao1998phase}, if the diffraction pattern is sufficiently oversampled, the phase can be retrieved from the diffraction pattern via iterative algorithms (see for instance \cite{ShechtmanEtAl2015Phase}).

Representing the object of interest by the complex-valued matrix $\mb X \in \Cp^{n \times n}$ and the captured diffraction pattern by $\mb Y \in \mathbb{R}^{m \times m}$, in the conventional phase retrieval configuration the diffraction pattern $\mb Y$ is obtained through the application of the following forward process:
\begin{align} \label{eq:fwd_process}
\mb Y = \abs{\mc F \paren{\mb X}}^2
\end{align}

Here, $\mc F$ represents the Fourier transform, converting the spatial information contained in the object matrix $\mb X$ into the frequency domain. $\abs{\cdot}$ is applied element-wise.
the original object  $\mb X$ is placed in an empty background such that $m = 2 \times n$. In simulation, this oversampling is achieved by zero-padding $\mb X$. The goal of the phase retrieval is to recover the image $\mb X$ from $\mb Y$.

Generally, some information about the support of the object (i.e. the location of the object) within the empty background is known. Let $\mb S_{0}\in \Cp^{m \times m}$ be the known support information. $S_{0}$ is a matrix containing ones where the object of interest is estimated to be, zeros otherwise. In general one may not know the exact support of the object of interest, but only an estimate. 

With these assumptions, the phase retrieval problem can be formulated as an optimization problem:
\begin{align}
    \min_{\hat{\mb X} \in \Cp^{n \times n}} \; \ell\paren{\mb Y, \abs{\mc F \paren{\hat{\mb X}}}^2}, \nonumber \\ 
    s.t. \paren{1-\mb S_{0}} \odot\hat{\mb X} = [\mb 0]_{m \times m} \label{eq:general_PR} 
\end{align}
where the first term imposes that \ref{eq:fwd_process} is satisfied (magnitude constraint), while the second term is the support constraint. 

\begin{figure*}    
  \centering
  \includegraphics[width=0.7\textwidth]{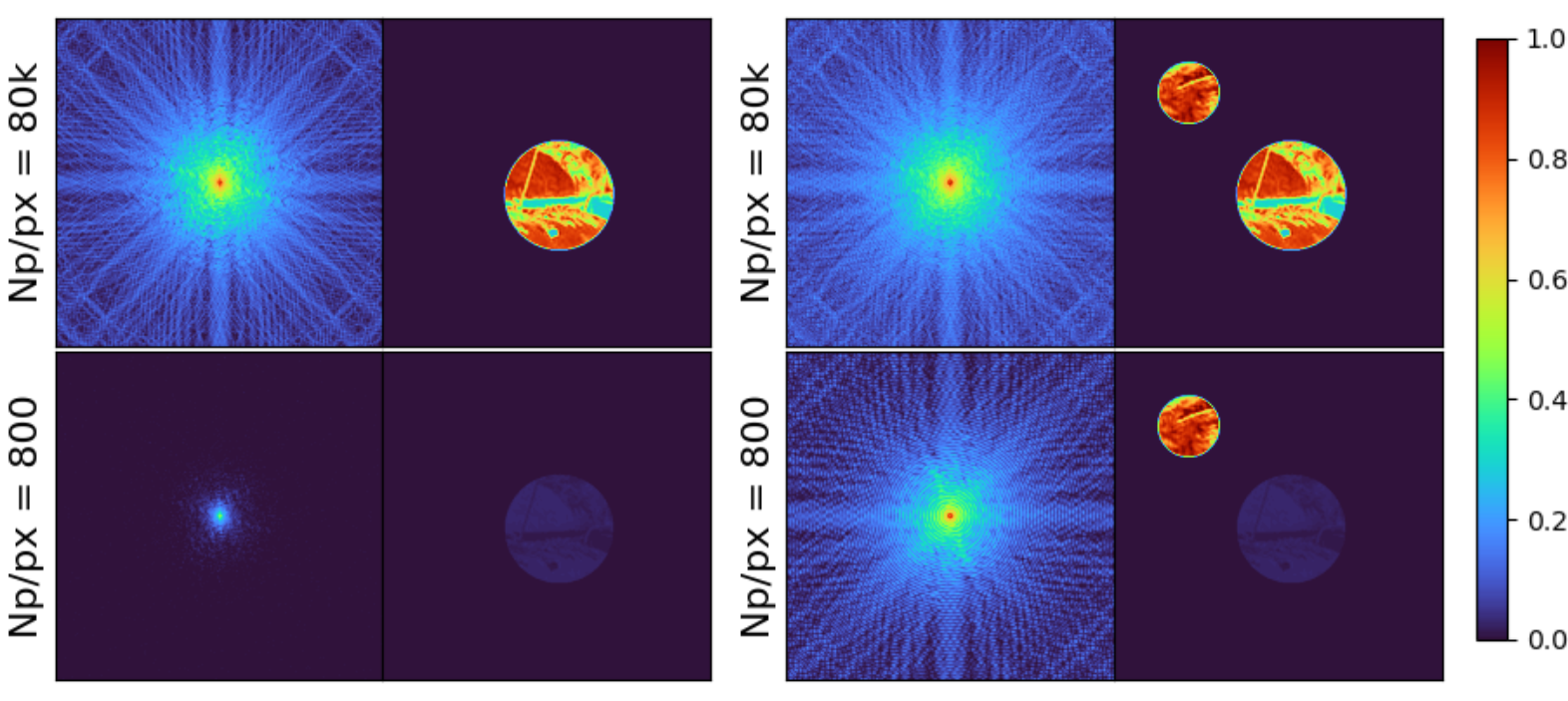}
  \caption{Examples from the simulated data at different levels of illumination of the sample region as measured in photons per pixel (Np/px). \textbf{[Left]} The first two columns are the diffraction pattern and the image sample without a static structure. \textbf{[Right]} And the last two columns are the same sample with a static region. \textbf{[Top]} The first row has a high illumination of 80k Np/px on both the sample and the static region. \textbf{[Bottom]} Whereas the second row has a high illumination of 80k Np/px on the static region and a low illumination of 800 Np/px on the sample. At lower illumination levels (bottom row), the sample exhibits significantly reduced pixel values. \annot{Lesser illumination results in higher presence of Poisson noise. Thus different illumination levels correspond to different levels of noise.}}  
  \label{fig:data_samples}
\end{figure*}

\subsection{Phase retrieval for low-light imaging}
\parahead{Challenge}
The major challenge of imaging at low-photon counts is the strong presence of noise in the captured signal. In practice, most of the existing PR methods \cite{fienup1978, miao2000oversampling, BauschkeEtAl2002Phase, luke2004, rodriguez2013, pham2019generalized} struggle to work in the low dose (high noise) setting. This comes in addition to the existing challenge of working without accurate support information which itself is sufficient for many of the existing algorithms to fail. \\

\parahead{Existing solutions}
Attempts have been made at low-photon phase retrieval that modify the imaging setup of the problem \cite{putkunz2011phase,lan2014method,lo2018situ}. 
While \cite{putkunz2011phase} relies on a phase diverse approach, \cite{lan2014method,lo2018situ} use in the imaging setup a static region made of heavy metals (usually gold) which can withstand high energy light.
Among these methods, in-situ CDI \cite{lo2018situ} achieves a substantial order of magnitude dose reduction over other methods. 
However, in-situ CDI uses a time-series of diffraction patterns which should contain a time-invariant "static" region. The static region's presence in each time step's measurement offers a strong overlap-in-time constraint for facilitating the convergence of phase retrieval optimization.
Fourier holography \cite{mcnulty1992high,barmherzig2019holographic} also utilizes two regions of interest. However, unlike Fourier holography, LoDIP can handle varying illuminations and doesn’t impose geometric constraints between the regions.

\subsection{Data-driven phase retrieval}
\annot{Data-driven methods have shown promise for phase retrieval \cite{SinhaEtAl2017Lensless, deng2020learning, UelwerEtAl2019Phase} including at low-dose \cite{GoyEtAl2018Low}. Subsequently, the end-to-end supervised learning paradigm has been extended by incorporating known physical models in the neural-network training. This includes utilizing an unrolled iterative algorithm \cite{wang2020phase,zhang2021physics} or by devising a self-supervised loss function \cite{boominathan2018phase,manekarend} capable of learning from measurements alone or using initial phase estimates from a physical model as inputs to the network \cite{rivenson2018phase,kang2020phase}. For a comprehensive review of data-driven phase retrieval, we refer the reader to \cite{wang2024use}.}

\annot{However, these methods are not yet widely adopted by microscopy practitioners due to several challenges: they often require large datasets for training, extensive computational resources, and their performance is heavily reliant on the quality of the training data. Additionally, retraining is necessary to adapt the model to different experimental settings. In this study, we propose a single-image deep learning method based on deep image prior that overcomes these limitations, requiring no large dataset and being adaptable across various experimental setups.}

\subsection{Zero-shot deep learning for phase retrieval}

\parahead{Pre-trained generative priors.}
Following the breakthroughs achieved by deep generative models in approximating image distributions \cite{kingma2013auto,goodfellow2020generative,kingma2018glow,chung2022diffusion}, such models have been leveraged as priors to regularize visual inverse problems \cite{bora2017compressed,chung2022diffusion}, including phase retrieval \cite{metzler2021deep}. Methods based on pre-trained generative priors are commonly referred to as ``zero-shot'' as no additional task-specific training is performed in these methods. Despite being unsupervised and not reliant on labeled datasets, these approaches still necessitate pretraining on extensive datasets. Additionally, akin to supervised learning methods, these pre-trained generative priors face challenges in generalization when the training data deviates from the target domain.

\parahead{Implicit generative priors.} A parallel line of work discovered that  even a convolutional neural network (CNN) without training can serve as a robust prior for natural images, and can be effectively employed to regularize inverse problems on such images. Beginning with the so-called deep image prior (DIP) \cite{ulyanov2018deep}. This has paved the way for a suite of methods including implicit neural representations (INR) \cite{mildenhall2021nerf,sitzmann2020implicit}, deep decoder \cite{heckel2018deep}, double-DIP \cite{gandelsman2019double}, random projection based methods \cite{li2023deep} and enhanced versions of DIP \cite{shi2022measuring}. For an exhaustive review, readers are directed to \cite{lu2022all}. In the absence of an overarching terminology for these methods, we label them as `implicit generative priors'. This term highlights the inductive bias derived from the network architecture rather than the training data. It also differentiates these methods from the use of pre-trained generative priors. Both types are frequently referred to as 'zero-shot’ deep learning methods. \annot{Finally, it highlights the generative nature of our algorithm which takes noise as input and generates a natural image.}

\parahead{For Phase Retrieval.} \annot{Several of these have also been applied to computational imaging inverse problems \cite{darestani2021accelerated,qayyum2022untrained,ongie2020deep} including simpler versions of phase retrieval \cite{jagatap2019algorithmic,heckel2018deep,bostan2020deep,tayal2021phase}. \footnote{\annot{Certain self-supervised data-driven methods have demonstrated efficacy even with a single sample \cite{boominathan2018phase, WangEtAl2020Deep}, resembling a zero-shot approach akin to an implicit generative prior. Instead of a random input seed, these methods rely on a sample-dependent input \cite{wang2024use}.}} However, the settings addressed in these works are considerably simpler compared to the single-shot far-field (Fraunhoffer) diffraction with high noise scenario considered in this paper. A naive application of DIP and related methods is known to struggle in such realistic phase retrieval scenarios \cite{tayal2021phase,zhuang2022practical}}.

%% file: method.tex
\begin{figure*}[t!]
  \centering
        \includegraphics[width=0.56\textwidth]{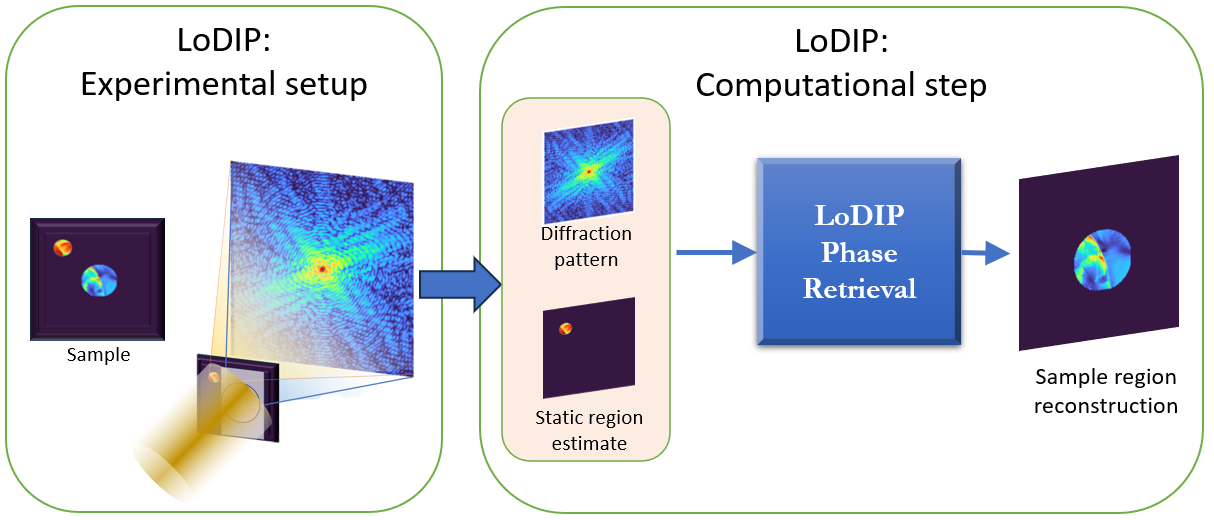}
        \includegraphics[width=0.4\textwidth]{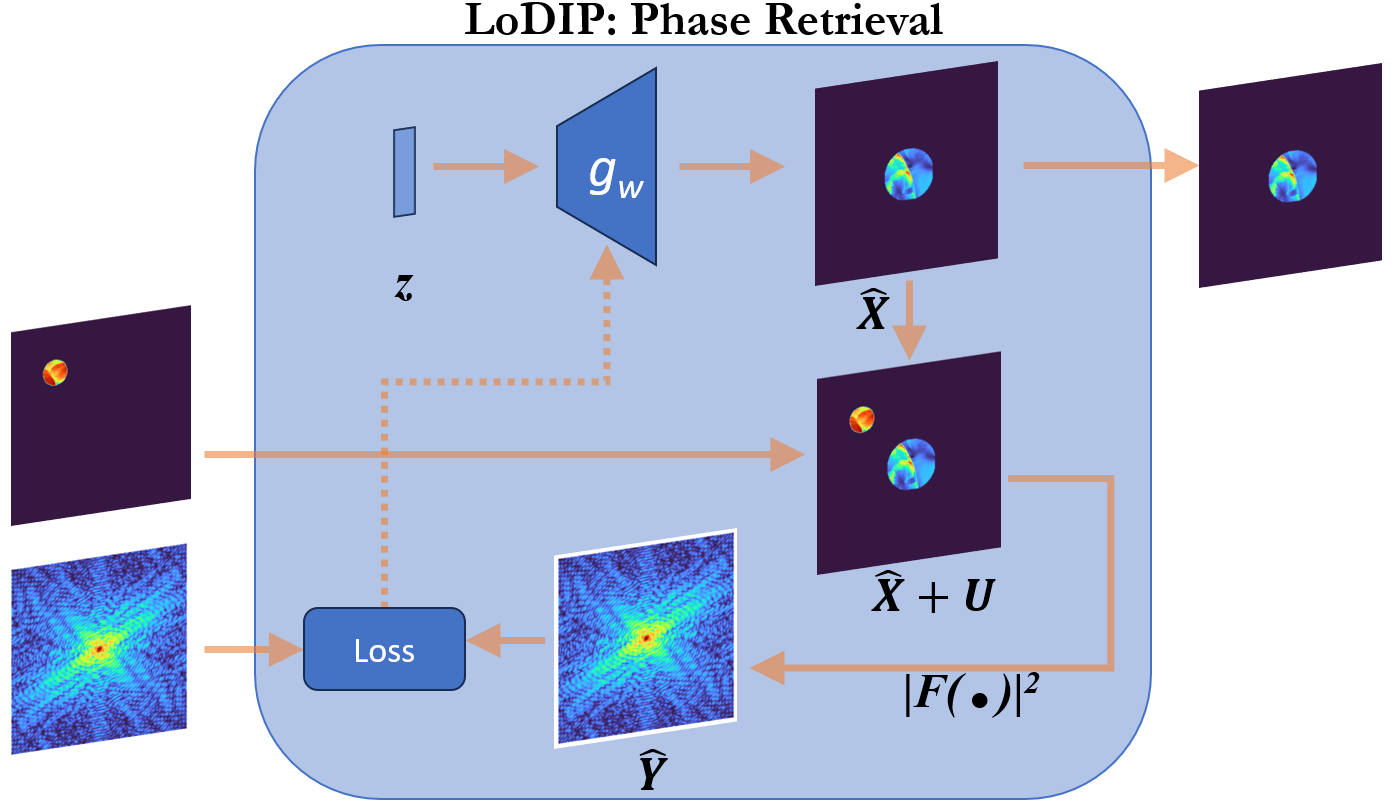}
  \caption{Coherent Diffraction Imaging (CDI) employs a coherent X-ray beam directed at a sample, capturing the resulting diffraction pattern on a 2D detector. A computational algorithm is then applied to reconstruct the desired sample image. \textbf{(Left)} \annot{Inspired by in-situ CDI \cite{lo2018situ}, LoDIP introduces two modifications to the CDI setup. First, it involves imaging the sample alongside a static region. Secondly, the static region is exposed to a high radiation dose, while the sample's exposure is significantly reduced. A customized sample grid and holder is used with an attenuator placed on top of the sample to reduce the incident radiation dose on the sample.} \textbf{(Center)} In the computational step, LoDIP takes both the diffraction pattern and an estimated reconstruction of the static region as inputs, generating a sample reconstruction as its output. \textbf{(Right)} LoDIP uses the output of an implicit generative model $g_{w}$ as an estimate of the sample $\hat{X}$ and iteratively updates the generator parameters $w$ to refine the estimate.}  
  \label{fig:lodip_schematic}
\end{figure*}

\begin{figure*}[ht]
  \centering
  \includegraphics[width=0.9\textwidth]{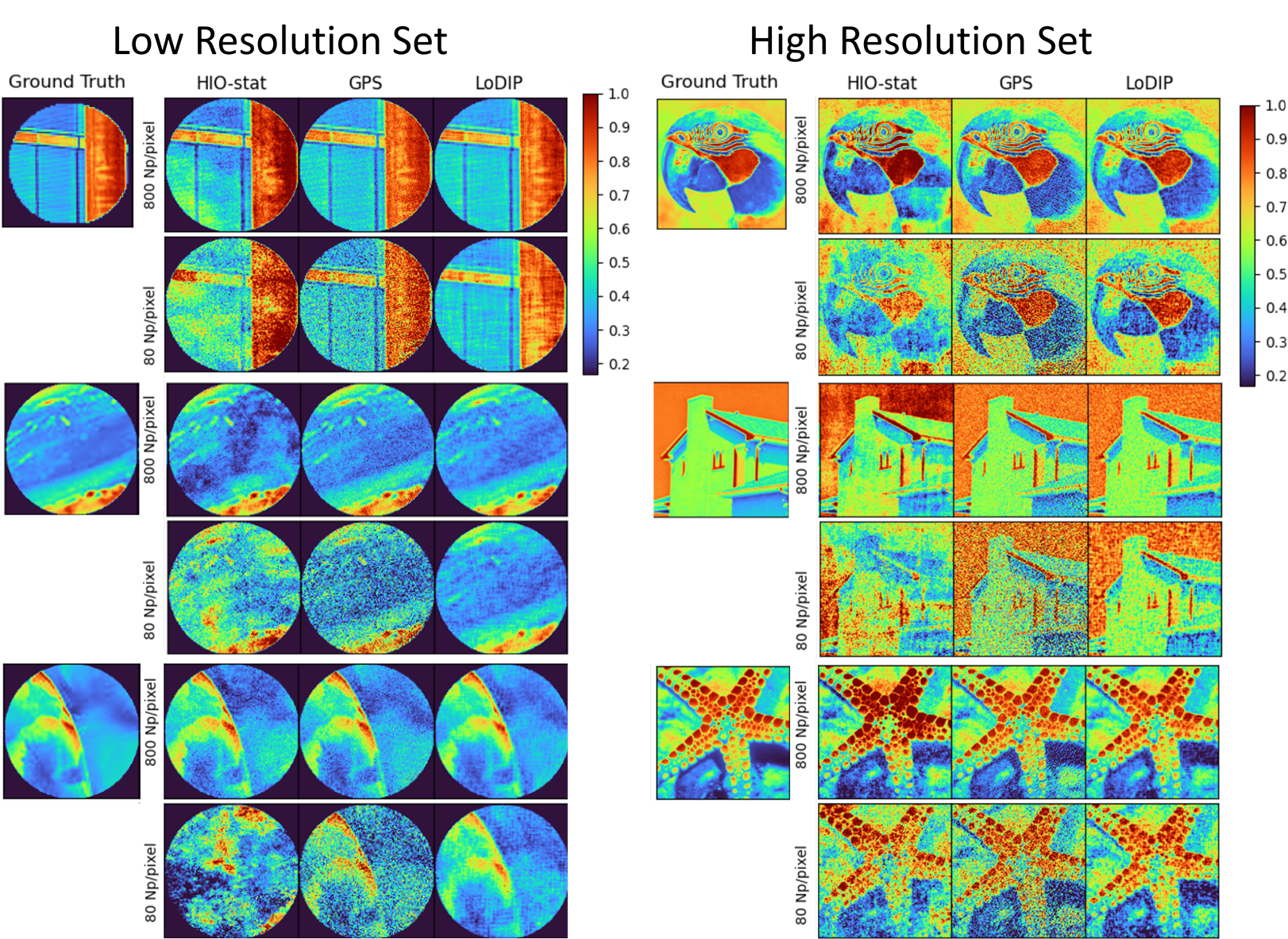}
  \caption{Experimental Results on simulated data. \textbf{Left:} Reconstruction of Low Resolution Set images for 800 Np/pixel (Top rows) and  80 Np/pixel (Bottom rows). \textbf{Right:} Reconstruction of High Resolution Set images for 800 Np/pixel (Top rows) and  80 Np/pixel (Bottom rows). Each image shows a zoomed-in view of only the sample region.}
  \label{fig:ch3_experimental_results}
\end{figure*}

\label{method}
The proposed method, which we refer to as LoDIP, modifies the data acquisition setup of conventional CDI to incorporate a high-dose static region inspired by the in-situ CDI setup. While in-situ CDI leverages the static region as a time-invariant constraint to reconstruct a sequence of measurements capturing a dynamic process, the high-dose static region serves a different purpose in LoDIP where we are interested in the single-image setting.

Firstly, it increases the available light for image formation on the detector, effectively reducing the impact of shot noise. Additionally, given the high-dose illumination, it is easy to produce an accurate reconstruction of the static region. The known static region is then used as a strong constraint to improve convergence of the phase retrieval optimization. Secondly, it mitigates the ambiguities arising from symmetries in the forward process which is a fundamental difficulty of phase retrieval, (see for instance \cite{ShechtmanEtAl2015Phase,manekar2020deep,hyder2020solving}). 

While these adaptations significantly enhance the performance of established phase retrieval methods (such as HIO) at low radiation doses, our experiments (see \cref{fig:ch3_experimental_results}), show that LoDIP can further improve the final reconstruction quality by incorporating an implicit generative prior for phase retrieval.

\subsection{LoDIP: experimental setup} \label{method_lodip_setup}

\annot{The sample of interest is placed within a finite support next to a static region of heavily scattering, dose-tolerant object such as a gold (Au) pattern on an optical stage. The X-ray illumination on the dose-sensitive sample is reduced to a tolerable limit by the presence of an attenuator mounted to the sample support (see Figure \ref{fig:lodip_schematic}), while the static region is exposed to the full dose of the incident illumination. The design of the customized sample support would need to include a SiN attenuator over the dose-sensitive region with a thickness tailored to the desired dose reduction factor and the energy of the incident x-ray beam. For example, to achieve a dose reduction factor of three orders of magnitude at a soft x-ray beamline operating within the water window as considered in our simulations ($\sim525 eV$), a \(Si_{3}N_{4}\) thickness of roughly 1.5 $\mu$m is required \footnote{This estimated value was obtained using the tabulated atomic scattering factors for solid \(Si_{3}N_{4}\) \cite{LBNLwebsite, HenkeScatterFactors1993}}}. 

Far-field diffraction patterns recorded from this setup are formed by the interference in Fourier space between the high-dose static region and the sample. These contain Poisson noise relative to the total illumination on the detector. Given the high-dose illumination on the static region and the known support, it is possible to obtain a high-quality static region reconstruction. In this work we use an established iterative method, Generalized Proximal Smoothing (GPS) initialized with 1000 iterations of HIO.

The placement of the object and the static structure do not influence the reconstruction given by LoDIP, as long as their supports do not overlap, i.e. $\mc S \paren{\mb X} \odot  \mc S \paren{\mb U}  = [\mb 0]_{m \times m}$, where $\mc S \paren{\cdot}$ extracts the support the image.

\subsection{LoDIP: Phase Retrieval}
For LoDIP, the known static region in the above data acquisition step provides us with a useful constraint to regularize and improve the convergence of the optimization problem. By incorporating the static structure $\mb U \in \Cp^{k \times k}$ in the original phase retrieval formulation, the forward process becomes:
\begin{align} 
    \mb Y = \abs{\mc F \paren{\mb X + \mb U}}^2 
    \label{eq:fwd_process_ref}
\end{align}
Here $\mb X$ and $\mb U$ are appropriately zero-padded to have the same dimension. In a similar way the optimization problem becomes:
\begin{align} 
    \min_{\mb X \in \Cp^{n \times n}} \; \ell\paren{\mb Y, \abs{\mc F \paren{\mb X + \mb U}}^2}, \nonumber\\
    s.t. \paren{1-\mb S_{0}} \odot \mb X = [\mb 0]_{m \times m}
\label{eq:general_PR_2}
\end{align}

In LoDIP this setup is further generalized by incorporating an \dip{}. For a given inverse problem aimed at recovering an image $\mb X \in \Cp^{n \times n}$ from its measurements $\mb Y \in \Cp^{m \times m}$ the \dip{} optmization problem is formulated as:
\begin{align} 
    \min_{\mb W} \; \ell\paren{\mb Y, \mc A \paren{g_{\mb W} \paren{\mb z}}}, 
\end{align}

where $\mb Y = \mc A \paren{\mb X} + \eta$, $\mc A$ is the known forward operator and $\eta$ is the noise. The optimization variable $\mb X$ is re-parameterized with a new function $g_{\mb W} \paren{\mb z}$, generally a CNN. Here $\mb W$ are the learnable parameters of the network and $\mb z$ is a random input seed which is fixed throughout the optimization process. $g_{\mb W}$ essentially works as a generative network which is optimized to output a single image.

Further, we extend this formulation to incorporate the constraints coming from the above setup: we include the known static region, the known sample support as well as the relative illumination doses on sample and static regions. 
Finally the optimization problem solved by LoDIP is:
\begin{align} 
    \min_{\mb W} \; \ell\paren{\mb Y, \abs{\mc F \paren{g_{\mb W} \paren{\mb z} + \mb U}}^2}, \nonumber\\
    s.t. \paren{1-\mb S_{0}} \odot\mb X = [\mb 0]_{m \times m}
\label{eq:general_PR_3}
\end{align}

\parahead{Extensions to LoDIP framework}
The LoDIP framework is highly versatile and can accommodate a variety of modifications. Firstly, it can incorporate any suitable \dip{} based architecture, including variants of DIP \cite{sun2021coil,chen2022unsupervised}. Secondly, LoDIP can be easily tailored to different experimental conditions by integrating them into the forward operator. This adaptability is demonstrated in Section \ref{results}, where experimental data captured using a probe was used. In this instance, LoDIP was adjusted to include the known probe function, in contrast to existing state-of-the-art methods like GPS \cite{pham2019generalized}, which would necessitate significant alterations for different experimental setups.

Moreover, LoDIP can smoothly integrate a rough reconstruction of the sample as initialization, enabling it to be effectively combined with other techniques aimed at generating high-quality initializations for iterative phase retrieval methods \cite{manekar2020deep}. Finally, LoDIP can be conveniently modified to function in the multiple measurements setting of the original in-situ CDI paper.


%% file: m_experiments.tex
\newcommand{\std}[1]{\annot{($\pm$#1})}

\begin{table*}[h]
    \centering
    \caption{Quantitative comparison of LoDIP and HIO-stat. For each metric, the reported values represent the mean over all the samples in the dataset. \annot{The standard deviation reported for PSNR shows the robustness of the method to different random inputs.}}
    \begin{tabular}{|c|c|c|c|c|c|c|}
    \hline
         & \multicolumn{3}{|c|}{Np/px=800}      & \multicolumn{3}{|c|}{Np/px=80} \\
    \hline
         & \multicolumn{6}{|c|}{Low resolution Set}\\
    \hline
         & \multicolumn{1}{|c|}{PSNR $\uparrow$} & \multicolumn{1}{c|}{SSIM $\uparrow$} & \multicolumn{1}{c|}{$R_{\text{real}}\downarrow$} & \multicolumn{1}{|c|}{PSNR $\uparrow$} & \multicolumn{1}{c|}{SSIM $\uparrow$} & \multicolumn{1}{c|}{$R_{\text{real}}\downarrow$} \\
    \hline

    LoDIP &         \textbf{28.52 \std{2.28}} & \textbf{0.76} & \textbf{0.07}  & \textbf{25.09 \std{1.18}} & \textbf{0.63} & \textbf{0.10}\\
    HIO-stat &      21.55 \std{0.77} & 0.47 & 0.18  & 13.11 \std{1.48} & 0.17 & 0.44\\
    GPS &           22.96 \std{0.20} & 0.53 & 0.13  & 13.91 \std{0.03} & 0.29 & 0.39\\    
    \hline
& \multicolumn{6}{|c|}{High Resolution Set}\\
    \hline
    LoDIP &         \textbf{23.56 \std{0.55}} & \textbf{0.59} & \textbf{0.09}  &  \textbf{20.01 \std{0.65}} & \textbf{0.42} & \textbf{0.14}\\
    HIO-stat &      19.90 \std{1.15} & 0.57 & 0.13  & 13.99 \std{1.10} & 0.24 & 0.29\\
    GPS &           19.16 \std{0.30} & 0.45 & 0.15  & 10.83 \std{0.04} & 0.17 & 0.40\\    
    \hline
    
         & \multicolumn{6}{|c|}{Biological cell sample}\\
    \hline
    LoDIP &         \textbf{26.97 \std{0.40}} & \textbf{0.68} & \textbf{0.11}  & \textbf{19.08 \std{0.99}} & \textbf{0.32} & \textbf{0.26}\\
    HIO-stat &      21.27 \std{1.44} & 0.50 & 0.18  & 12.21 \std{1.62} & 0.14 & 0.39\\
    GPS &           23.68 \std{0.31} & 0.53 & 0.15  & 15.14 \std{0.04} & 0.19 & 0.42\\    
    \hline
    
    \end{tabular}    
    \label{tab:isdip_testset_error}
\end{table*}
\subsection{Data} \label{data_sim}
We perform experiments on three kinds of data. First, we create a simulated data\footnote{Data is available at https://github.com/raunakmanekar/LoDIP-PR.} using a procedure similar to \citep{chang2023deep}. Natural images are used to create the sample region and the static region for each sample in the dataset. Examples of the generated images can be seen in \cref{fig:data_samples}. \annot{Next, using the data provided in \cite{lo2018situ} we create a simulated image of biological cell using physically accurate simulations of a gold lacey for the static structure and a biological cell for the sample region,these can be found in \cref{fig:ch3_experimental_results_bio_sample}, \ref{fig:loose_supp}.}

Finally, we demonstrate the applicability of LoDIP on experimental diffraction patterns
from live glioblastoma cells (see Fig. \ref{fig:real_data_exp}). 
\annot{In this setup, the collimated beam with a wavelength of 534 $nm$ incident onto a pair of 100 \(\mu m\) pinholes with an edge-to-edge separation distance of 100 \(\mu m\). This produced a pair of beams incident onto the sample 400 \(\mu m\) downstream of the dual pinhole aperture with the same intensity, as no dose reduction was performed in this experiment. The diffraction intensity from the sample exit wave was recorded using a CCD detector placed at the focal point of a lens adjacent to the sample.}
The static structure is a 100 {\textmu} pinhole exposed to the same incident illumination as the sample. Unlike the first two datasets above, this data has been collected using a probe.  This dataset is sourced from Lo et al. \cite{lo2018situ}. Further information about the optical laser data collection can be found in the original paper. 

\paragraph{Simulated data generation}
The experimental setup is similar to the one in \cite{lo2018situ}. The size of the entire image (here, 512x512) corresponds to the size of the detector in the real experimental setup. 
The support of the sample region is 170x170 pixels. This gives an oversampling ratio of approximately 3.0 which is higher than the theoretical requirement of 2. The sample and the static structure have non-overlapping support and the static structure is set to have half the radius of the sample region (however, the method is independent of the static region size). 

\paragraph{Simulating low-light conditions} Based on previous studies on low-dose phase retrieval \cite{lo2018situ} and \cite{lu2023computational}, the illumination on the sample has been varied from 2.5x$10^5$ to 2.5x$10^9$ photons per ~{\textmu}m$^{-2}$. The illumination on the static structure has been fixed at 2.5x$10^{9}$ photons per ~{\textmu}m$^{-2}$. We show results on the low doses of 2.5x$10^6$ and 2.5x$10^7$ photons per ~{\textmu}m$^{-2}$ which corresponds to $80$ and $800$ photons per pixel (Np/px) based on the simulation geometry.

\subsection{Results}\label{results}
\parahead{Methods} For all experiments we compare the relative performance of the proposed method (LoDIP) with other popular methods which can be used in this setup. Specifically, we compare with Hybrid input-output (HIO) \cite{fienup1982phase};  Deep Image Prior \cite{ulyanov2018deep} (DIP) with no static region information; HIO-stat which is a modification of HIO that uses a static region; and Generalized proximal smoothing (GPS) \cite{pham2019generalized} which is the state of the art method for phase retrieval. HIO and DIP work with diffraction patterns collected without a static region (see column 1 in \cref{fig:data_samples}. Rest of the methods, namely HIO-stat, GPS and LoDIP use the static region, however LoDIP also exploits neural network priors.

\begin{figure*}[ht!]
  \centering
  \includegraphics[width=1.0\textwidth]{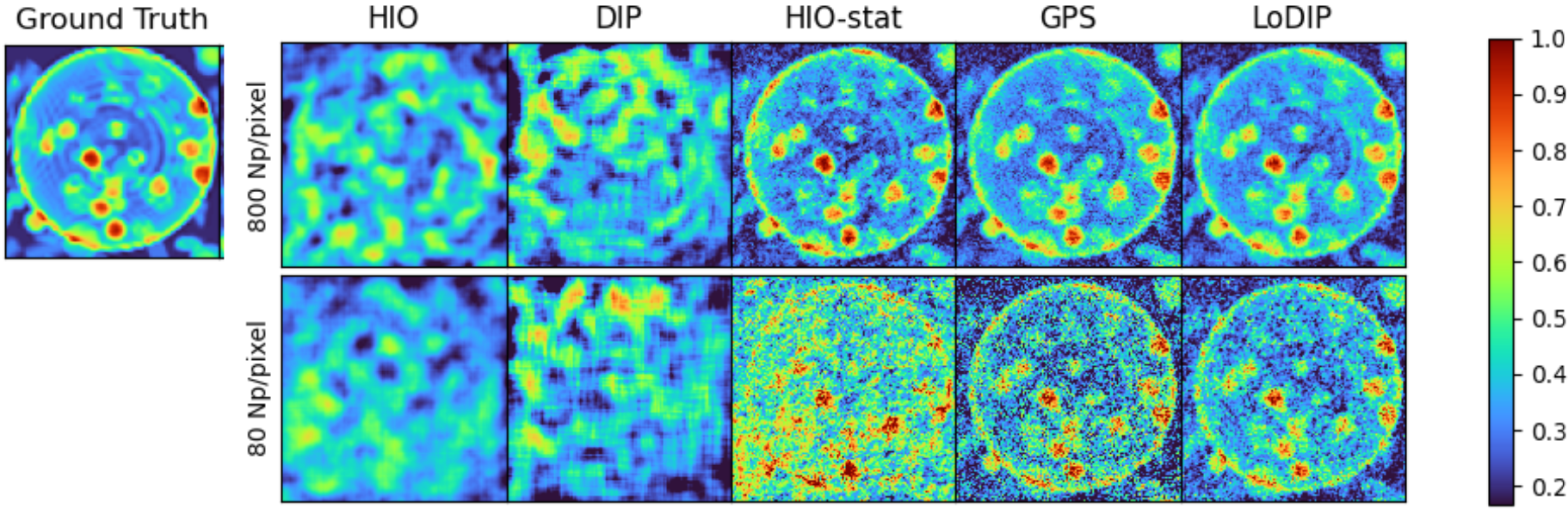}
  \caption{Experimental Results on biological cell sample. \textbf{(Top row)} Reconstruction at 800 Np/pixel. \textbf{(Bottom row)} Reconstruction at 80 Np/pixel).}  
  \label{fig:ch3_experimental_results_bio_sample}
\end{figure*}

\begin{figure}[ht!]
  \centering
  \includegraphics[width=0.4\textwidth]{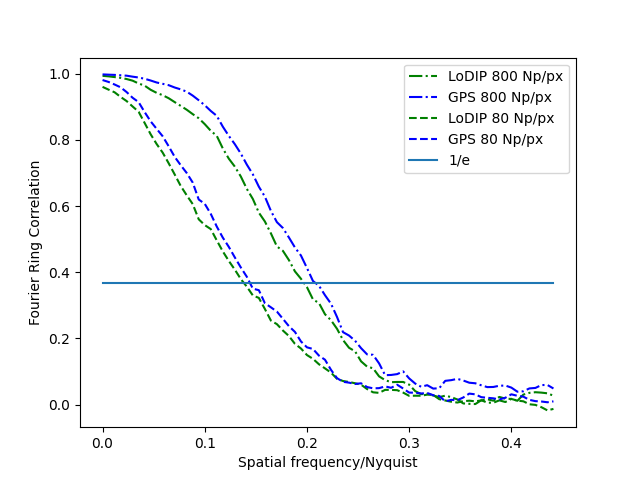}
  \caption{Comparison of FRC values for GPS and LoDIP reconstructions on biological cell at 800 Np/px and 80 Np/px.}
  \label{fig:Frc_bio}
\end{figure}

\parahead{Metrics.} Following previous works on reconstruction with noisy measurements \cite{metzler2018prdeep}, we measure the reconstruction accuracy for all experiments in this paper using using Peak Signal-to-Noise Ratio (PSNR). Higher values of PSNR indicate better fidelity; Structural Similarity Index (SSIM) \cite{lawrence2021low} which measures the similarity between two images, evaluating luminance, contrast, and structure for enhanced accuracy in assessing perceptual image quality (higher the better); R-factor ($R_{real}$) \cite{pham2019generalized} which measures the degree of agreement between observed and predicted data, commonly used in the assessment of image reconstruction quality (smaller the better).

\paragraph{Reconstruction of natural images}

Two commonly used natural image datasets for testing phase retrieval algorithms were selected: the Set12 dataset \cite{metzler2018prdeep,zhang2017beyond,lawrence2021low} and internet-sourced stock images used in \cite{chang2023deep}. For each image, diffraction patterns were first generated using the forward model outlined in \cref{method} and \cref{data_sim}. Results were presented for two low-dose scenarios: 800 and 80 photons per pixel. \annot{
Different doses correspond in practice to different amounts of Poisson noise. Good reconstruction across dose amounts shows the robustness of the method with respect to different noise levels.} 

We observed that the Set12 images showed higher resolution and finer details compared to the stock images, hence we distinguish them as the high and low resolution sets, respectively. The complementary characteristics of these datasets allow for a thorough evaluation of each method's denoising capability and achievable resolution. Across each dataset, we compared reconstructions from all methods using 12 samples, each utilizing a different image for its static structure and sample region.

\cref{tab:isdip_testset_error} shows results averaged over the entire test set of 12 images for the low and high resolutions sets. Since DIP and HIO without reference regions did not provide good results in this case we omit those methods here. LoDIP attains better results across all metrics for both data sets and photon counts. The better performance of LoDIP is especially clear in the low photon counts case where LoDIP accuracy is approximately double the accuracy of the other methods. Finally, as expected, the error is higher for the high resolution set as more fine detail has to be reconstructed. 

\parahead{Discussion} \cref{fig:ch3_experimental_results} displays a comparison of the results of LoDIP, HIO-stat and GPS for images in the low resolution set (left) and high resolution set (right). To study the robustness of our method with respect to the experimental setup we used a circular mask for low resolution set and a square one for the high resolution set. GPS and LoDIP yield the best results, however, GPS generates noisier images compared to LoDIP, highlighting LoDIP's strong denoising ability. The figure shows also the choice of the mask for the sample region does not influence the reconstruction quality for LoDIP (and for the other methods).

We also note that iterative methods such as HIO-stat and GPS require parameter tuning by experts in the field and often multiple independent runs of the algorithm are necessary to obtain good reconstruction. On the other hand, LoDIP hyper-parameter tuning is minimal and convergence is usually obtained in one run. 

\paragraph{Reconstruction of biological cell sample}
Computational microscopy represents a major potential application of the proposed method. Following previous works on PR for CDI \cite{lo2018situ,lu2023computational}, we evaluated the performance of LoDIP on realistic and physically accurate simulated data for a prototypical live cell. The static structure and biological cell were simulated as done in \cite{lo2018situ}. The results in this section use a simulated 20-nm thick gold lacey pattern as the static structure and a simulated cell consisting of a vesicle containing water and protein aggregates. The static region is unknown and estimated using the procedure described in \cref{method_lodip_setup}.

\parahead{Discussion} 
Similar to the case of natural images, \cref{fig:ch3_experimental_results_bio_sample} shows that GPS and LoDIP produce the best reconstructions in both cases with LoDIP providing a less noisy reconstructed image than GPS especially at 80 Np/px. HIO and DIP without reference regions fail in the reconstruction. The last section of \cref{tab:isdip_testset_error} shows that LoDIP outperforms the other methods across all metrics.

\parahead{Comparing Resolution.} 
\cref{fig:Frc_bio} displays the Fourier Ring Correlation (FRC) for both photon counts cases for LoDIP and GPS. Higher FRC values correspond to better resolution. The resolution of the reconstruction obtained by LoDIP is comparable to the one obtained by GPS. This indicates that the denoising capability of LoDIP results in only a minimal loss of fine-scale details.

\begin{figure*} [ht]
  \centering
  \includegraphics[width=0.8\textwidth]{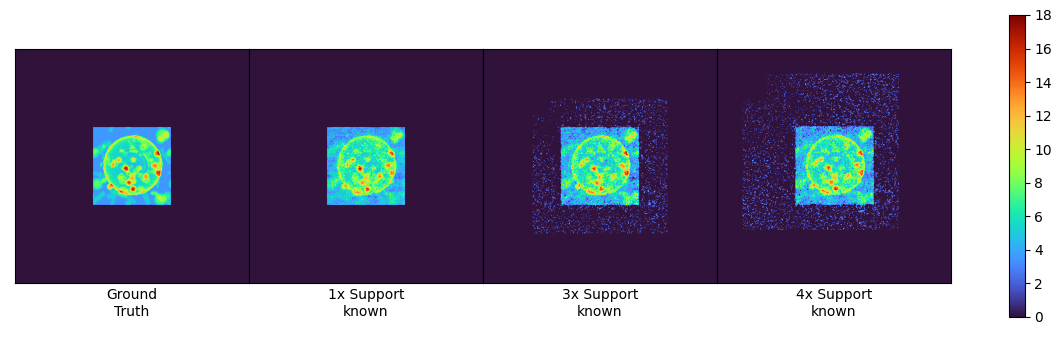}
  \caption{Reconstruction using LoDIP at 80 Np/px without knowledge of accurate support. The columns from left to right are (1) The ground truth, (2) reconstruction with knowledge of the exact support (PSNR=20.08), (3) reconstruction with an approximate support of 3x the actual size (PSNR=20.06), (4) 4x the actual size (PSNR=19.85). Both visually and quantitatively, the LoDIP reconstruction is robust to inaccurate specification of the support. PSNR is calculated only over the exact sample region.}  
  \label{fig:loose_supp}
\end{figure*}

\parahead{Reconstruction without precise support}
PR without precise support information presents a common and challenging problem.
From \cref{fig:loose_supp}, we can see that LoDIP reconstructions are robust to inaccurate support specification. Widely used phase retrieval algorithms struggle when accurate support information is not available \cite{tayal2021phase,zhuang2022practical} and require additional measures for support estimation\cite{marchesini2003x}.

\begin{figure*}[ht!]
  \centering
  \includegraphics[width=0.7\textwidth]{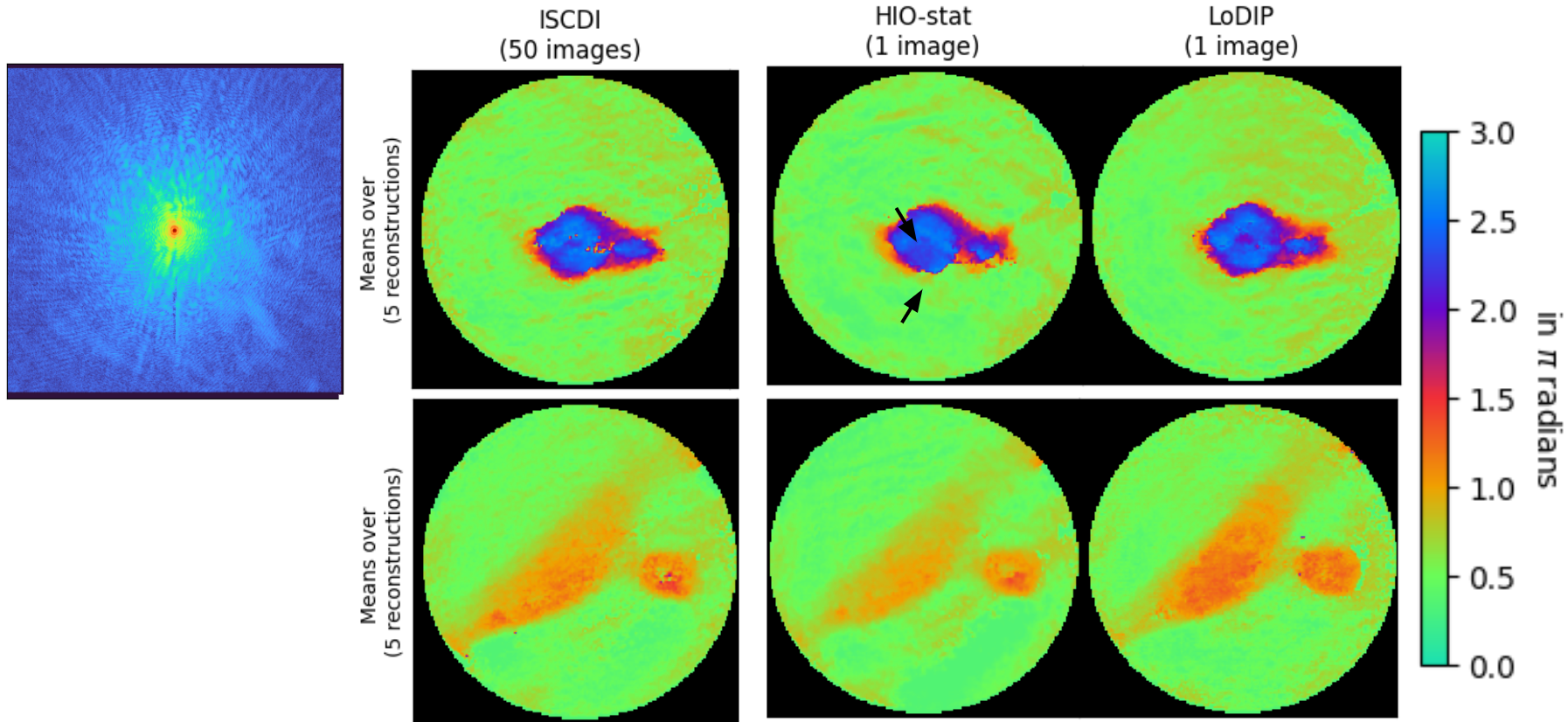}
  \caption{Reconstruction of experimental data \textbf{(Left)} An experimentally captured diffraction pattern. \textbf{(Right)} Reconstruction of two samples from different methods. The mean of the top five reconstructions from 20 independent runs is shown in the figure. The static and sample images in this example have the same high dose.  LoDIP is especially advantageous in the low dose setting, as the previous experiments show, so as expected, in this high dose setting LoDIP obtains comparable results as in-situ CDI and HIO in terms of $R_F$. However, LoDIP uses a single diffraction pattern to produce a reconstruction of comparable quality as in-situ CDI which uses 50 samples. Moreover, while HIO-stat reconstruction has similar $R_F$ as LoDIP, it contains visible artefacts (indicated by the arrows).}
  \label{fig:real_data_exp}
\end{figure*}

\paragraph{Reconstruction from experimental data} 
Finally, we demonstrate the proposed method on experimentally captured diffraction patterns (one of them is shown in \cref{fig:real_data_exp} left column). Unlike the previous experiments, the object is complex-valued and the optical setup includes a probe. We note explicitly that in this dataset the static and sample images have the same high dose, so this is not exactly the setup for which our method is designed. In fact, our method is especially advantageous in the low dose setting, as the previous experiments show. 
Nonetheless, this setup is very close to the one of interest so we included it in this work. 

\parahead{Comparison with in-situ CDI.} The original in-situ CDI method \cite{lo2018situ} uses 50 diffraction patterns with a fixed time-invariant static structure required in all the images. HIO-stat and LoDIP use only a single diffraction pattern and perform single-image phase retrieval. This is an advantage over in-situ CDI since as in many experimental setups a truly static structure is difficult to ensure because of the data collection procedure (e.g. in collecting tomographic data, which requires shifting or tilting both the sample and the static structure). In this section, the reconstruction given by in-situ CDI is used as a proxy for the ground truth. In fact, a comparison between in-situ CDI and LoDIP would be unfair since LoDIP uses 50 times less data (only one diffraction pattern instead of 50).

\parahead{Comparison with GPS.} A comparison with GPS is not possible for this data since the current version of GPS does not support Fresnel propagation and adding this would require major modifications which are outside of the scope of this work.
In these experiments, LoDIP and HIO-stat have been modified to incorporate the known probe function.

\parahead{Metrics.} Since there is no ground truth available, to evaluate our results we use R-factor in the Fourier domain. This measures the disagreement between the captured diffraction pattern $\mb Y$ to the Fourier magnitudes $\lvert\mc F \paren{\hat{\mb X}}\rvert$ of the reconstruction $\hat{\mb X}$ and is defined as:

\begin{gather} \label{eq:r_factor}
    R_F(\hat{\mb X}) = \frac{\sum_{i,j}\abs{\lvert\mc F \paren{\hat{\mb X}}\rvert_{i,j} - \mb Y_{i,j}}}{\sum_{i,j} Y_{i,j}}
\end{gather}

The reconstruction obtained through in-situ CDI yields an R-factor of $30.22\%(\pm0.97\%)$. As explained before in-situ CDI utilizes 50 diffraction patterns for its reconstruction. Comparing this R-factor directly with LoDIP, which only uses a single diffraction pattern, would be unfair due to the substantial difference in data quantity needed to obtain the reconstruction. However, we can use in-situ CDI R-factor as a benchmark to assess the performance of our method. 

\parahead{Discussion.} The average $R_F$ from 20 independent reconstructions based on a single diffraction pattern using both LoDIP and HIO-stat is presented in \cref{tab:real_data_rf}. Remarkably, LoDIP achieves a comparable R-factor to in-situ CDI, highlighting its efficacy even without the need for multiple diffraction patterns, making it less data-intensive. Interestingly, HIO-stat also demonstrates a comparable R-factor to LoDIP in this scenario. It's important to note, however, that this success is primarily attributed to the high photon count setting of this experiment, where many methods perform well. It is crucial also to note that, as demonstrated in previous experiments, HIO-stat's performance diminishes in low photon count settings, providing less accurate results than LoDIP in this more realistic case. Finally, \cref{fig:real_data_exp} shows the averaged top 5 reconstructions out of 20 independent runs for all methods. We see that HIO-stat introduces visible artifacts in its reconstruction while LoDIP does not (again the in-situ CDI reconstruction is treated as a proxy for the ground truth in this case). Therefore, these results highlight the robustness and adaptability of LoDIP, showcasing its ability to easily accommodate changes in the data acquisition setup and maintain reliable reconstructions across varying experimental conditions.

\begin{table}[ht!]
    \centering
    \caption{Quantitative comparison on experimental data. The mean and standard deviation of R-factor($R_F$) values are calculated over 20 independent reconstructions each of 5 different samples.}
    \begin{tabular}{ll}
    \hline
         & $R_F$ \\
    \hline
    HIO-stat &          34.70\%($\pm$0.03\%) \\
    LoDIP &           33.20\%($\pm$0.03\%) \\
    \hline
    \end{tabular}    
    \label{tab:real_data_rf}
\end{table}